\documentclass[12pt]{iopart}
\usepackage{graphicx}
\usepackage{subfig}
\usepackage{iopams}
\usepackage{cite}
\usepackage{xcolor}
\usepackage{stackengine}

\begin{document}

\def\stackalignment{l}

\title{Radiation induced acceleration of ions}
\author{E G Gelfer$^{1}$, A M Fedotov$^2$ and S Weber$^{1,3}$ }
\address{$^1$ELI-Beamlines, Institute of Physics of the Czech Academy of Sciences, Dolni Brezany 252 41, Czech Republic}
\address{$^2$National Research Nuclear University MEPhI (Moscow Engineering Physics Institute), Moscow 115409, Russia}
\address{$^3$School of Science, Xi'an Jiaotong University, Xi'an 710049, China}
\ead{egelfer@gmail.com}

\begin{abstract}
Radiation friction can have a substantial impact on electron dynamics in a transparent target exposed to a strong laser pulse. In particular, by modifying quiver electron motion, it can strongly enhance the longitudinal charge separation field, thus stimulating ion acceleration. We present a model and simulation results for such a radiation induced ion acceleration and study the scalings of the maximal attainable and average ion energies with respect to the laser and target parameters. We also compare the performance of this mechanism to the conventional ones.
\end{abstract}


\submitto{\NJP}

\section{Introduction}
Several new generation multi petawatt (PW) laser facilities are under construction in Europe  \cite{ELI,ELINP,Apollon}, while even more powerful $100-200$ PW projects have been announced \cite{100PW,XCELS}. A promising application of these powerful lasers is ion acceleration in a plasma. In contrast to laser driven electron acceleration, for which a great progress is observed even with the existing lasers and reaching a $8$ GeV energy level has been recently reported \cite{elacc78}, energies of laser accelerated ions in experiments have not yet attained the level of $100$ MeV \cite{100mev1,100mev2}. Though it is sufficient for such applications as proton radiography  \cite{radiography,borghesi} or materials stress testing \cite{stress}, some others, including hadron therapy for cancer treatment  \cite{therapy, therapy2} or fast ignition of nuclear fusion  \cite{naumova2009}, require higher ion energy \cite{borghesi, daido, macchi}.

A number of mechanisms for laser driven ion acceleration in a plasma are known. In the context of high power lasers the radiation pressure acceleration \cite{esirkepov2004,macchi2009,qiao2009,bulanov2010,bulanov2015,bulanov2016,shen2017,wilks1992,naumova2009,schlegel2009,psikal2018}  (RPA) stands out by a strong (linear) scaling of the ion energy with laser intensity. Irrespectively to that the details are peculiar to the cases of thin  \cite{esirkepov2004,macchi2009,qiao2009,bulanov2010,bulanov2015,bulanov2016,shen2017} and thick \cite{wilks1992,naumova2009,schlegel2009,psikal2018} targets (these cases are called the light sail (LS) and hole boring (HB) regimes, respectively; target thickness should be compared to the skin depth \cite{bulanov2016}), the main idea of RPA is that a strong laser pulse, being reflected off a dense plasma target, pushes the plasma electrons forward, thus creating a charge separation longitudinal electric field, which in turn accelerates the ions of the target. 

Furthermore, the prospective high power facilities mentioned at the beginning are designed to generate laser pulses in the intensity range  $10^{23}-10^{24}$ W/cm$^2$. In such a strong field electrons radiate so violently that their motion will be strongly influenced by the recoil  \cite{RDR2004,dipiazza2012,burton2014}. This effect, called radiation friction (RF), was recently studied in the specially dedicated experiments on head-on collision of intense laser pulses with energetic electrons  \cite{cole2018,poder2018} and ultrarelativistic positrons propagating in silicon \cite{wistisen2018}. In particular, it was clearly  demonstrated \cite{cole2018,poder2018} that the recoil changed both, the electron energy distribution and the Compton radiation spectrum, hence it should be properly taken into account in an accurate theoretical interpretation of the experimental data. 

In spite of the notable activity, the current understanding of the impact of RF on ion acceleration remains fragmentary. On the one hand, it was demonstrated that RF has a limited impact on ion acceleration in the LS regime \cite{tamburini2010,tamburini2011,tamburini2012}. This is natural, as in an opaque target only a minority of electrons may appear in a strong field region probing the RF force. In contrast, it looks as though RF can noticeably reduce ions energy in the HB case  \cite{capdessus2015}. Furthermore, RF can substantially modify the laser-plasma dynamics in a transparent target \cite{vshivkov1998,gelfer2019}. In particular, it was established analytically and confirmed numerically \cite{pw_scirep,pw_ppcf} that RF can enhance charge separation in a transparent laser driven plasma. Though it was also demonstrated by Particle-in-cell (PIC) simulations that RF can enhance ion acceleration under such conditions \cite{chen2011,capdessus2015}, no convincing explanation was proposed so far.

Classically, radiation recoil is described by introducing a RF force $\mathbf{F}_{\mathrm{RF}}$ into the electron equation of motion
\begin{equation}\label{lleq}
\frac{d\mathbf{p}}{dt}=-e\left(\mathbf{E}+\frac1{c}[\mathbf{v}\times\mathbf{B}]\right)+\mathbf{F}_{\mathrm{RF}}.
\end{equation}
The full-length expression for the force $\mathbf{F}_{\mathrm{RF}}$ is derived and discussed in \cite{LL}. However, as we prove below (see Appendix A), under the conditions of our interest it is enough to retain only the leading term
\begin{equation}\label{rfforce}
\mathbf{F}_{\mathrm{RF}}\approx-\frac{2}{3}\frac{\mathbf{v}}{c}\alpha eE_{cr}\chi^2,
\end{equation}
directed oppositely to the electron velocity. Here $e>0$ and $m$ are electron charge magnitude and mass, $\alpha=e^2/\hbar c$ is the fine structure constant, $E_{cr}=m^2c^3/e\hbar$ is the critical field of quantum electrodynamics, and
\begin{equation}\label{chi}
\chi=\frac{e\hbar}{m^2c^3}\gamma\sqrt{\left(\mathbf{E}+\frac{1}{c}[\mathbf{v}\times\mathbf{B}]\right)^2-\frac{1}{c^2}(\mathbf{Ev})^2},\quad \gamma=\sqrt{1+(\mathbf{p}/mc)^2}
\end{equation}
is called the dynamical quantum parameter \cite{ritus1985}. It is worth to stress that the expression (\ref{rfforce}) is of entirely classical origin. Though we intentionally write it here in terms of the quantum parameter $\chi$ to stress the significance of the latter in discriminating the classical regime of RF \cite{ritus1985}, the Planck constant ultimately cancels. 

Now consider a strong plane electromagnetic (laser) wave propagating in a plasma. As shown in \cite{pw_scirep}, the longitudinal component of the equation of motion for a plasma electron in such a field can be cast into the form 
\begin{equation}\label{eq1p}
\frac{du_x}{dt}=\frac{1}{2\gamma}\frac{d a^2(\varphi)}{d\varphi}+\mu a^4\frac{1-v_x}{1+v_x}-\sigma,\quad 
\mu=\frac{2}{3}\alpha\frac{\hbar\omega}{mc^2}\simeq 1.18\cdot 10^{-8} \frac{1}{\lambda[\mu m]}.
\end{equation}
Here $\mathbf{u}=\mathbf{p}/mc$ is dimensionless momentum of the electron, time $t$ is measured in units of $\omega^{-1}$, where $\omega=2\pi c/\lambda$ and $\lambda$ are the laser frequency and wavelength, electron velocity $\mathbf{v}$ is measured in the units of speed of light $c$, and $\sigma$ is the dimensionless (measured in units $mc\omega/e$) longitudinal charge separation field. Unless stated otherwise, we assume that the laser field is circularly polarized (CP), so that its dimensionless amplitude can be expressed as $\boldsymbol{a}_\bot\equiv e\mathbf{E}/m\omega c=\{0,a(\varphi)\cos\varphi,a(\varphi)\sin\varphi\}$, where $\varphi=\omega(t-x/c)$ is the phase and $a(\varphi)$ is the wave envelope. According to \cite{pw_scirep}, the validity of equation~(\ref{eq1p}) for the cold transparent plasma in the specified setup is restricted only by the assumptions that the propagating wave is of ultrarelativistic intensity ($a_0\gg1$) and the transverse component of the RF force is much smaller than the transverse Lorentz force ($u_x\gg \mu a_0^4$). 

\begin{figure}[ht!]
\begin{center}
\includegraphics[width=0.95\linewidth]{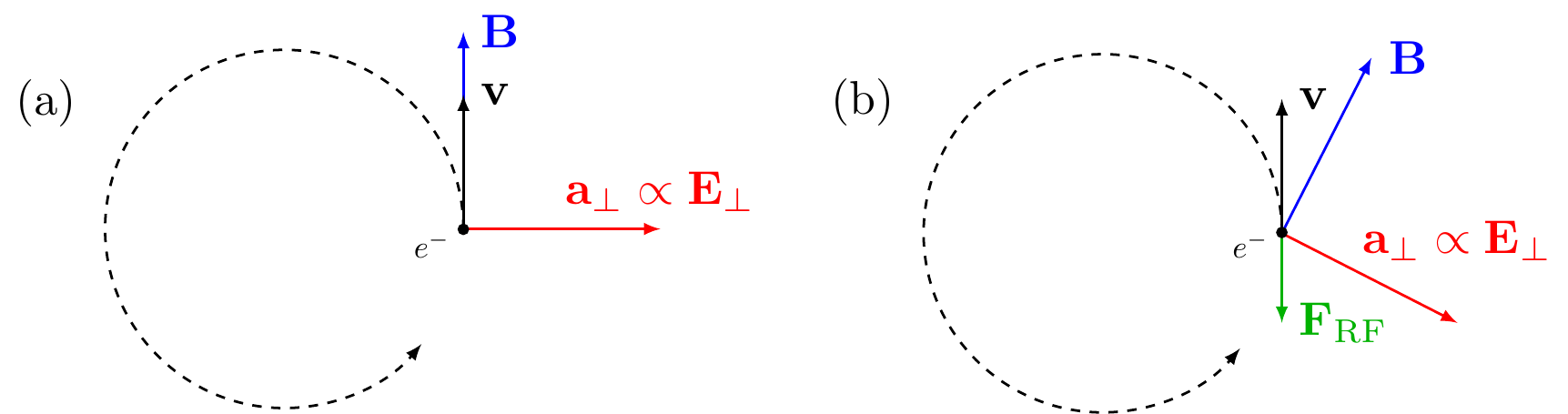}
\end{center}
\caption{\label{figexpl} The transverse projection of electron trajectory (dashed) and the field strengths $\mathbf{E}_\perp$, $\mathbf{B}$ of a CP plane wave at the instance when the electron velocity is directed upwards without (a) and with (b) account for RF.}
\end{figure}

The first term on the right hand side of equation~(\ref{eq1p}), proportional to the slope of the wave envelope, is the relativistic ponderomotive force \cite{bauer_prl1995}. The second term arises indirectly due to RF as expained below \cite{voronin1964,zeldovich1975,dipiazza2008,pw_scirep,pw_ppcf}. 
Notably it is positive, meaning that in presence of RF the electrons are pushed forward stronger. As a consequence, the longitudinal charge separation field in the plasma is enhanced. To understand such an apparently counter-intuitive behavior, assume that the wave envelope is flat ($a(\varphi)=\mathrm{const}$), then the ponderomotive force vanishes. If, furthermore, RF is neglected then the electron trajectory in the reference frame with the electron on average at rest is a circle in a plane transverse to the laser propagation direction, see Figure~\ref{figexpl} (a). In this case the velocity is parallel to the magnetic field, hence the Lorentz force $-e[\mathbf{v\times B}]$ in  Equation~(\ref{lleq}) is equal to zero and the accelerating longitudinal force does not appear. On the other hand, when taken into account, RF modifies the {\it transverse} electron motion by inducing an angle between the velocity and magnetic field as in Figure~\ref{figexpl} (b). The resulting {\it longitudinal} component of the {\it Lorentz force} accelerates the electron forward (out of the page of the drawing). It is worth to emphasize that the longitudinal component of RF force is directed opposite to $v_x$, i.e. backwards (as expected for a friction force), still the {\it resultant} of the longitudinal Lorenz and RF forces turns out to be positive.

Previously \cite{pw_scirep,pw_ppcf} we investigated, how this effect enhances the generation of longitudinal waves in an extended plasma slab, when the ions are so heavy, that their motion can be neglected (see \cite{pw_ppcf} for details). Here, on the contrary, we show how it can be applied for ion acceleration. We consider an intense laser pulse, incident normally on a thin transparent foil, with RF taken into account, and develop a 1D model capable for estimating the energy of accelerated ions for given laser and target parameters. We justify our model by PIC simulations and demonstrate that such radiation induced acceleration (RIA) mechanism can produce ions with considerably higher energy than RPA. 



In the present paper we consider a laser pulse that is strong enough to completely separate charges in the target. This regime is quite similar to a directed Coloumb explosion (DCE) considered as a valuable alternative mechanism for ion acceleration \cite{therapy,esirkepov2002,fourkal2005,ssbulanov2008,grech2009,grech2010}. However, in our case the ions are accelerated via attraction to the electrons dragged by laser pulse, whereas in DCE the lighter ions are accelerated via repulsion of heavier ones in a structured target with the laser pulse serving only to remove most of the electrons from the target.

\section{Analytical model 
}

On impinging on a target, a laser pulse of ultrarelativistic intensity pushes the electrons forward, instantly accelerating them almost to the speed of light. Snapshots from a typical 1D PIC simulation of a pulse incidence on an underdense plasma foil, with and without RF taken into account, are presented in Figure~\ref{fig1} (the details of the numerical approach are given in the Appendix B). One can see that with RF taken into account much more electrons are captured accelerating inside the laser pulse. As a consequence, the longitudinal charge separation field
is also substantially stronger than in Figure~\ref{fig1}(b).

\begin{figure}[ht!]
\topinset{(a)}{\subfloat{\includegraphics[width=0.5\linewidth]{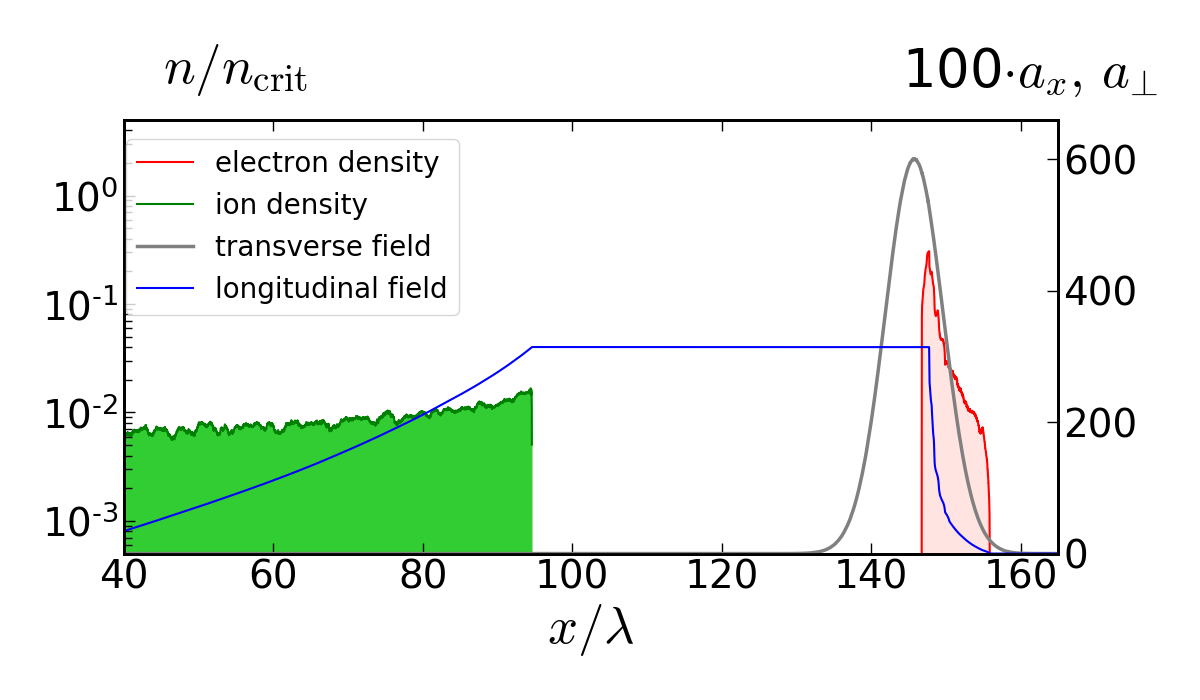}}}{0.9cm}{3.2cm}
\topinset{(b)}{\subfloat{\includegraphics[width=0.5\linewidth]{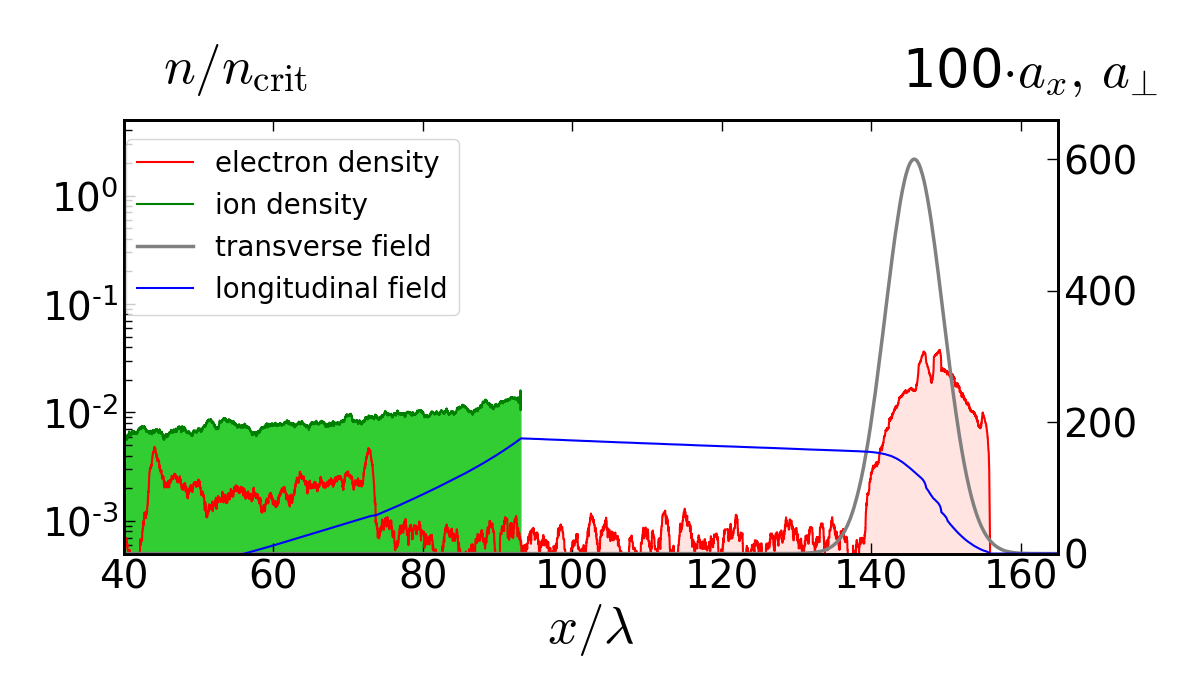}}}{1.23cm}{3.2cm}
\caption{\label{fig1} Particle densities and field distributions from 1D PIC simulations with (a) and without (b) RF at $t=600$ fs. Laser parameters: field strength $a_0=600$ (corresponding to the laser intensity $I\approx 10^{24}$ W/cm$^2$), FWHM pulse duration $30$ fs, circular polarization. Target parameters: density $n=0.5 n_c$ and thickness $d=\lambda=1\mu$m.}
\end{figure}

In order to study the ion motion, let us assume that the target is initially so thin, that for a time being the majority of electrons is moving inside the laser pulse, remaining completely separated from the ions accelerating behind the pulse (see Figure~\ref{fig1}) and generalize accordingly the model (\ref{eq1p}). First, since the spatial distribution of electrons is essential for the  ion acceleration, it is more convenient to rewrite the Equation (\ref{eq1p})  in the Eulerian form and assume that $u=u(t,x)$ and $d/dt=\partial/\partial t+v_x\partial/\partial x$, where spatial variable $x$ is dimensionalized by $c/\omega$. Second, the longitudinal charge separation field $\sigma(t,x)$ due to initial neutrality of the target inside the pulse coincides in 1D with the areal density of the charge, located after $x$ in the laser propagation direction, 
\begin{equation}
\sigma(t,x)=\int\limits_x^\infty \tilde{n}(t,x')dx',
\end{equation}
where $\tilde{n}=n/n_c$ and $n_c=m_e\omega^2/4\pi e^2$ is the plasma critical density.

Assume that the longitudinal electron motion is ultrarelativistic, $\gamma\approx u_x\gg u_\bot\approx a$, $1+v_x\approx 2$ and $1-v_x\approx a^2/2u_x^2$ . Then, using more convenient variables $(t,\varphi)$ instead of $(t,x)$, we have
\begin{equation}\label{model_rel}
\frac{\partial u_x}{\partial t}-\frac{\partial}{\partial\varphi}\left(\frac{a^2}{2u_x}\right)-\frac{\mu a^6}{4u_x^2}+\sigma=0.
\end{equation}
With the same notations and assumptions the continuity equation takes the form
\begin{equation}
\frac{\partial \sigma}{\partial t}+\frac{\partial \sigma}{\partial\varphi}\frac{a^2}{2u_x^2}=0.\label{cont_eq1}
\end{equation}

Under the above assumptions ion acceleration is governed by the total electron areal density inside the laser pulse
\begin{equation}
\sigma_p(t)=\sigma(t,\varphi=T)=\int\limits_{-\infty}^T \tilde{n}(t,\varphi')d\varphi',
\end{equation}
where $T=\omega t_{pulse}$ is the dimensionless pulse duration. Indeed, within our model, electrons after leaving the laser pulse mix with ions and don't influence ion acceleration anymore (this assumption is revised for the RIA case in the next section). Then the distribution of the dimensionless charge separation field behind the pulse at the moment $t$ can be written as.  
\begin{equation}\label{ax}
a_x(t,x)\sim\sigma_p(t)-\int\limits_x^{x_R} \tilde{n}_i(t,x') dx',
\end{equation}
where $x_R$ is the position of the rightmost ion. It takes the maximum value $\sigma_p(t)$ in $x_R$, and then vanishes inside the ion cloud, see Figure~\ref{fig1}. In particular, the maximal longitudinal momentum and energy of the ions (i.e. of the ions at $x_R$) are expressed by

\begin{equation}\label{emax}
u_{i,\rm max}(t)=\frac{m_e}{m_i}\int\limits_0^t \sigma_p(t')dt', \quad \mathcal{E}_{i,\rm max}=m_ic^2\,\sqrt{1+u_{i,\rm max}^2},
\end{equation}
where $m_i$ is the ion mass. 

To find the accelerating field $\sigma_p(t)$, let us consider separately the two concurring mechanisms of charge separation in a transparent thin foil, RIA, when the charge separation is governed by the third (RF induced) term in (\ref{model_rel}),  and Ponderomotive acceleration (PA), when it is governed by the second (ponderomotive) term  \cite{pw_scirep,pw_ppcf}.

\begin{figure}[ht!]
\begin{center}
\includegraphics[width=0.7\linewidth]{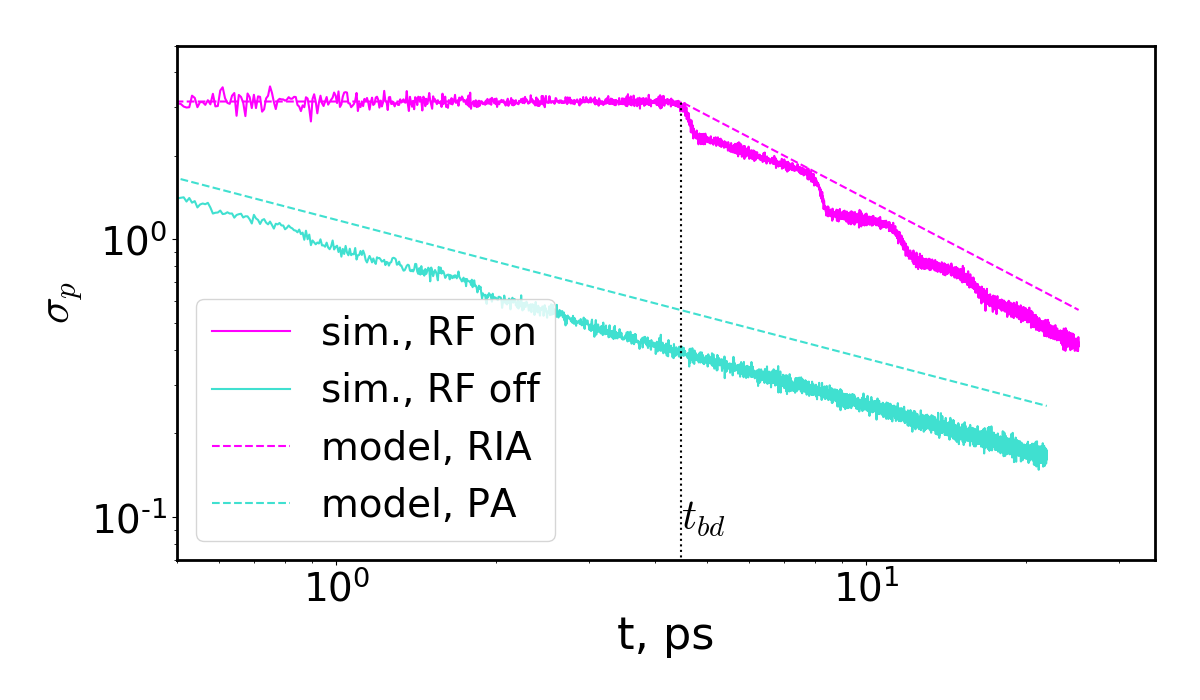}
\caption{\label{fig_sigma} The total charge of electrons, captured inside the laser pulse vs time. Solid lines -- 1D PIC simulations, dashed lines -- estimations (\ref{sigmaria}) and (\ref{sigmapa}).  Laser and target parameters are similar to Fig.~\ref{fig1}. }
\end{center}
\end{figure}

Figure~\ref{fig_sigma} shows that, with RF taken into account, the acceleration process splits into two stages. On the initial stage the charge captured inside the pulse is conserved, meaning that all the electrons from the target are moving inside the laser pulse. However, since they are slower than the pulse, they gradually drift from its front to the back. Eventually, at some moment $t_{\rm bd}$ when the electrons reach the region where the laser field is not strong enough to balance the electrostatic field, a breakdown occurs \cite{pw_scirep, pw_ppcf}. On the second stage the electron population leaks away from the back of the pulse, and the ion acceleration saturates. 

In the RIA case we neglect the second term in (\ref{model_rel}) and estimate the longitudinal 4-velocity of electrons inside the pulse by balancing the terms corresponding to the electrostatic force and the RF induced longitudinal Lorentz force:
\begin{equation}
u_x\sim \frac{a^3}{2}\sqrt{\frac{\mu}{\sigma}}\label{urf}.
\end{equation}

Let $\sigma_0=\omega dn_0/n_c$ be the dimensionless areal density of the target. Here $n_0$ and $d$ are its initial density and thickness. For $t<t_{\rm bd}$ the value  of the longitudinal field at the position of the leftmost electron inside the pulse is equal to $\sigma_0$. Until the breakdown the leftmost electron propagates through the pulse and acquires the phase $\varphi=T$ \cite{pw_scirep}. Therefore the breakdown time can be obtained from the condition 
\begin{equation}
\varphi(t_{\rm bd})=\int\limits_0^{t_{\rm bd}}(1-v_x)dt\sim T.
\end{equation}
In the ultrarelativistic case under consideration, $u_x\gg u_\bot\sim a$, and the longitudinal velocity of electrons can be estimated as $v_x\approx 1- a^2/2u_x^2$, where $u_x$ corresponds to (\ref{urf}) (for the of the analysis of the assumptions, made during the current derivation, see the Appendix A). Inside the pulse (i.e. while the field is strong enough to balance the charge separation  force with RF induced longitudinal force) we use $a\sim a_0$, where $a_0=a(0)$ is the amplitude of the laser field, and get
\begin{equation}\label{taubd}
t_{\rm bd}\approx \frac{\mu a_0^4 T}{2\sigma_0}.
\end{equation}
The meaning of this condition is precisely that a breakdown occurs when the leftmost electrons have passed to the back of the pulse.

For $t>t_{\rm bd}$ the charge inside the pulse can be calculated by incorporating (\ref{urf}) into the continuity equation (\ref{cont_eq1}).
After separating the variables [$\sigma(t,\varphi)=f(t)g(\varphi)$] we obtain the solution
\begin{equation}
\sigma\approx \frac{\mu}{2 t}\int a^4 d\varphi.
\end{equation}
Estimating $\int\limits_{-\infty}^T a^4 d\varphi\sim a_0^4 T$, we arrive at the expression for a charge captured inside the laser pulse 
\begin{equation}\label{sigmaria}
\sigma_p^{\rm (RIA)}(t)\sim\cases{\sigma_0,\quad &$t<t_{\rm bd}$\\ \mu a_0^4 T/2t,\quad &$t>t_{\rm bd}$\\}.
\end{equation}


In the PA case we neglect the third term in (\ref{model_rel}) and assume that $a(\varphi)=a_0\alpha(\varphi/T)$, where the function $\alpha(\zeta)$ is such that: (i) it rapidly vanishes for $|\zeta|\gtrsim1$, and (ii) $\alpha(0)=1$. By seeking a solution of the resulting equations in the form $u_x=a_0\sqrt{\tau/T}\alpha^2(\varphi/T)/g(\varphi/T)$, $\sigma=a_0\nu(\varphi/T)/(2\sqrt{T\tau})$, they transform to
\begin{equation}\label{p1}
\nu=g'-\frac{\alpha^2}{g},\quad 
\nu' g^2-\alpha^2\nu=0.
\end{equation}
Since they no more contain large or small parameters, even without solving them we can claim that for an arbitrary pulse shape the charge captured inside the pulse can be estimated up to a numerical factor as 
\begin{equation}\label{sigmapa}
\sigma_p^{\rm (PA)}(t)\equiv\sigma(t,\varphi=T)\sim\frac{a_0}{2\sqrt{Tt}},\quad t>t_0=\frac{a_0^2}{4T\sigma_0^2},
\end{equation}
where the moment $t_0$, starting from which the estimation (\ref{sigmapa}) becomes valid, is determined from the condition $\sigma_p^{\rm (PA)}(t_0)=\sigma_0$. 

Estimates (\ref{sigmaria}) and (\ref{sigmapa}) are in a good agreement with simulations results, see Figure~\ref{fig_sigma}. 

To justify the consistency of our analysis let us inspect the parameter range for which either one of the second or the third terms in the right hand side of (\ref{model_rel}) can be neglected. Using the same estimation $da^2/d\varphi\sim a_0^2/T$ as above, we can see that the RF induced term dominates over the ponderomotive term if
\begin{equation}\label{rfdom}
\mu\sigma_0a_0^2T^2>1.
\end{equation}

\section{Results}
\begin{figure}[ht!]
\topinset{(a)}{\subfloat{\includegraphics[width=0.5\linewidth]{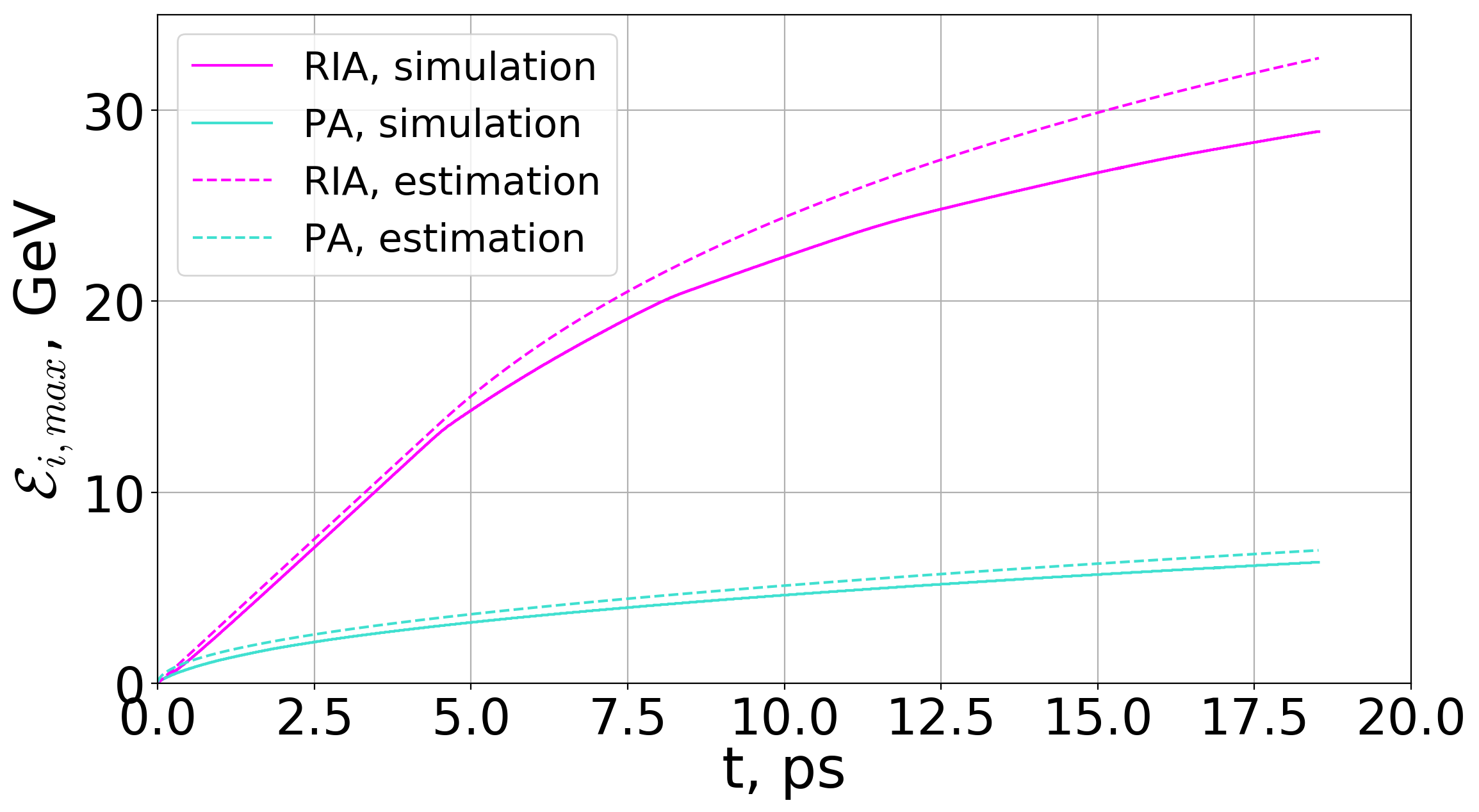}}}{0.2cm}{3.5cm}
\topinset{(b)}{\subfloat{\includegraphics[width=0.5\linewidth]{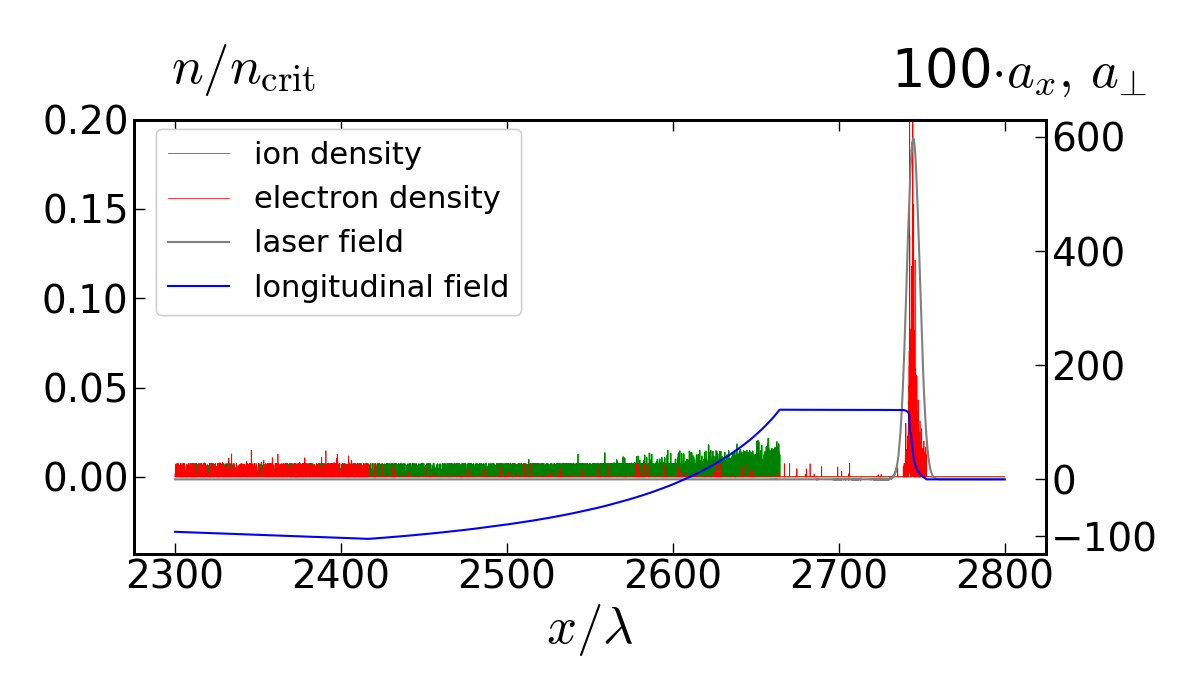}}}{1.2cm}{3.5cm}
\caption{\label{fig_maxen} (a): The dependence of maximal ion energy on time for different acceleration mechanisms; solid lines -- PIC simulations, dashed lines -- estimations (\ref{umaxria}), (\ref{umaxpa}). (b): Density and field distributions at $t=2t_{\rm bd}$ with RF taken into account. $a_0=600$ ($I\approx 10^{24}$ W/cm$^2$), CP, FWHM $30$ fs, target density $n=0.5 n_c$, thickness $d=\lambda=1\mu m$.}
\end{figure}

 According to the developed model, the time dependence of the maximal energy of the ions for RIA can be estimated by substituting the value of the electron areal density inside the laser pulse (\ref{sigmaria}) into (\ref{emax})
\begin{equation}\label{umaxria}
 \mathcal{E}^{\rm (RIA)}_{i,\rm max}(t)\sim m_ec^2\cases{
  \sigma_0t, &$t<t_{\rm bd}$,\\
 \frac{\mu a_0^4 T}{2}\left(1+\ln\frac{t}{t_{\rm bd}}\right), &$t>t_{\rm bd}$,\\}
\end{equation}
where for simplicity we assume that ions are ultrarelativistic, and $\mathcal{E}_{i,\rm max}\approx u_{i,\rm max}$. In the PA case, using (\ref{sigmapa}) we obtain
\begin{equation}
   \mathcal{E}^{\rm (PA)}_{i,\rm max}(t)\sim m_ec^2 a_0\sqrt{\frac{t}{T}},\label{umaxpa}
\end{equation}
Estimates (\ref{umaxria}) and (\ref{umaxpa}) are in a good agreement with 1D PIC simulations, see Fig.~\ref{fig_maxen}~(a). 

Let us now estimate the average energy of ions $\left<\mathcal{E}_i(t)\right>$, which provides more meaningful description of the acceleration process. 
According to (\ref{ax}), longitudinal charge separation field is positive for $x<x_R$, such that $\int\limits_x^{x_R} n_i(t,x') dx'<\sigma_p(t)$, and hence the areal density of ions, accelerating at the moment $t$, can be estimated as $\sigma_i(t)\sim\sigma_p(t)$. 
Then the increase of energy of all ions per cross section $S$ at the moment $t$ can be written as $d\mathcal{E}_i(t)/S=\sigma_p(t)\frac{\sigma_p(t)}{2}dt$, where we assumed that the accelerating field $a_x(t,x)$ averaged over $x$ is equal to $\sigma_p(t)/2$. 

Hence the average energy of ions can be estimated as
\begin{equation}\label{eav}
\left<\mathcal{E}_i(t)\right>\sim\int\frac{\sigma_p^2(t)}{2\sigma_0} dt,
\end{equation}
Substituting (\ref{sigmaria}) and (\ref{sigmapa}) to (\ref{eav}) we get
\begin{equation}
 \left<\mathcal{E}^{(\rm RIA)}_i(t)\right>\sim \cases{
  m_ec^2\frac{\sigma_0t}{2}, &$t<t_{\rm bd}$,\\
 m_ec^2\frac{\mu a_0^4 T}{4}\left(2-\frac{t_{\rm bd}}{t}\right), &$t>t_{\rm bd}$\\}\label{eavria}
\end{equation}
and
\begin{equation}
    \left<\mathcal{E}^{(\rm PA)}_i(t)\right>\sim m_ec^2 \frac{a_0^2}{8\sigma_0 T}\ln\frac{t}{t_0}.\label{eavpa}
\end{equation}

Let us briefly discuss the estimation (\ref{eavria}) for the RIA case. First, we note, that for $t<t_{\rm bd}$, $\left<\mathcal{E}^{(\rm RIA)}_i(t)\right>\sim\mathcal{E}^{\rm (RIA)}_{i,\rm max}(t)/2$, and it is natural, since all the particles participate in acceleration, and the accelerating field is distributed from its maximum value $\sigma_0$ to zero. However for large $t$ the estimate (\ref{eavria}) is not accurate enough, because after the breakdown electrons are accelerating leftwards by strong longitudinal field, and can create negative longitudinal field, which decelerates ions, see Figure~\ref{fig_maxen}(b). The estimate can be refined, if one takes into account that the acceleration stops, when half of the electrons leave the laser pulse, propagate through the ion cloud and balance the quasistatic accelerating field of the electrons, remaining inside the pulse, i.e. at $t=2t_{bd}$. Then the final averaged energy of ions can be estimated as 
\begin{equation}\label{eavlim}
 \left<\mathcal{E}^{(\rm RIA)}_i\right>_{\rm lim}\sim m_e c^2\frac{3\mu a_0^4 T}{8},\quad t>2t_{\rm bd}.
\end{equation}



\begin{figure}[ht!]
\topinset{(a)}{\subfloat{\includegraphics[width=0.5\linewidth]{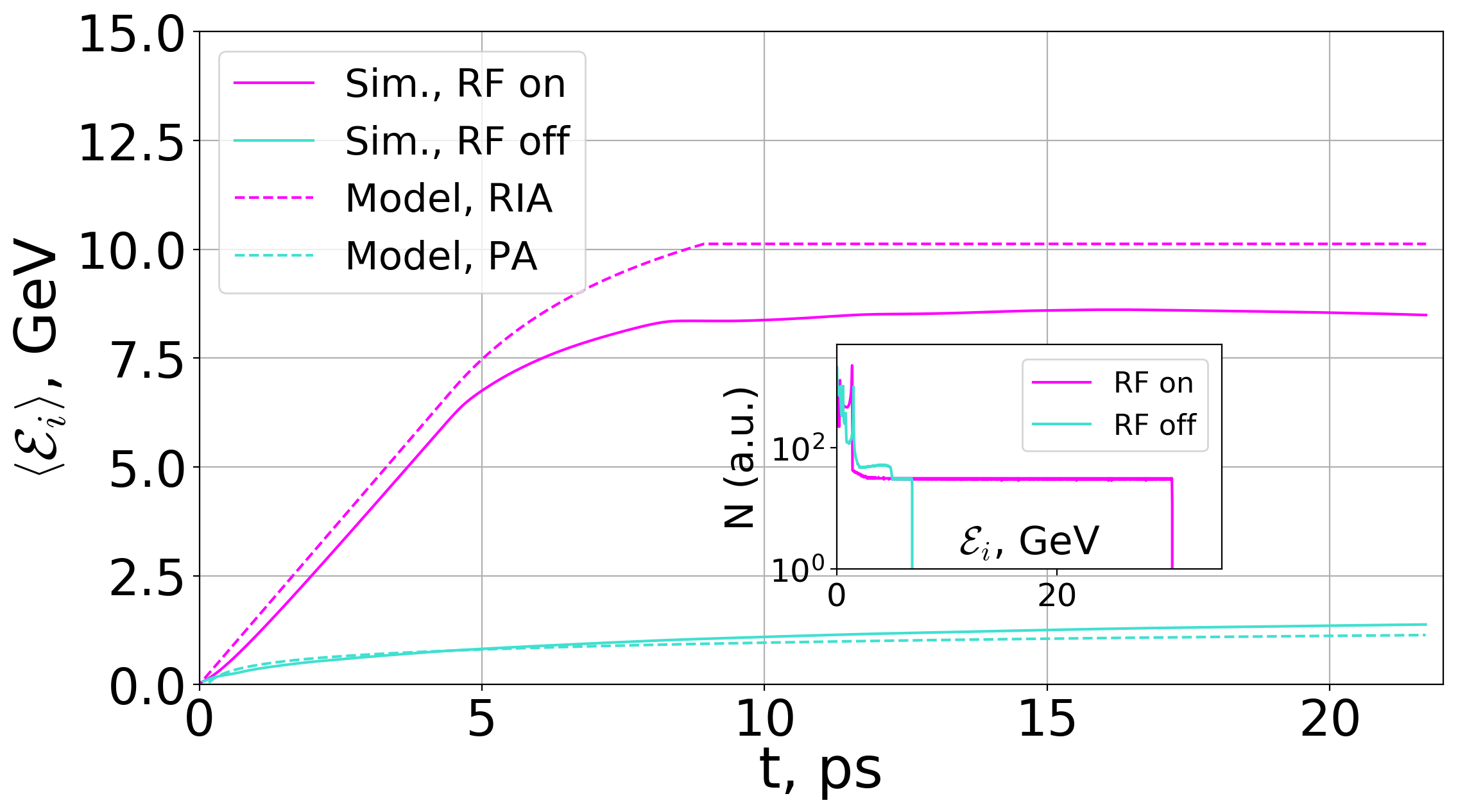}}}{0.25cm}{3.5cm}
\topinset{(b)}{\subfloat{\includegraphics[width=0.5\linewidth]{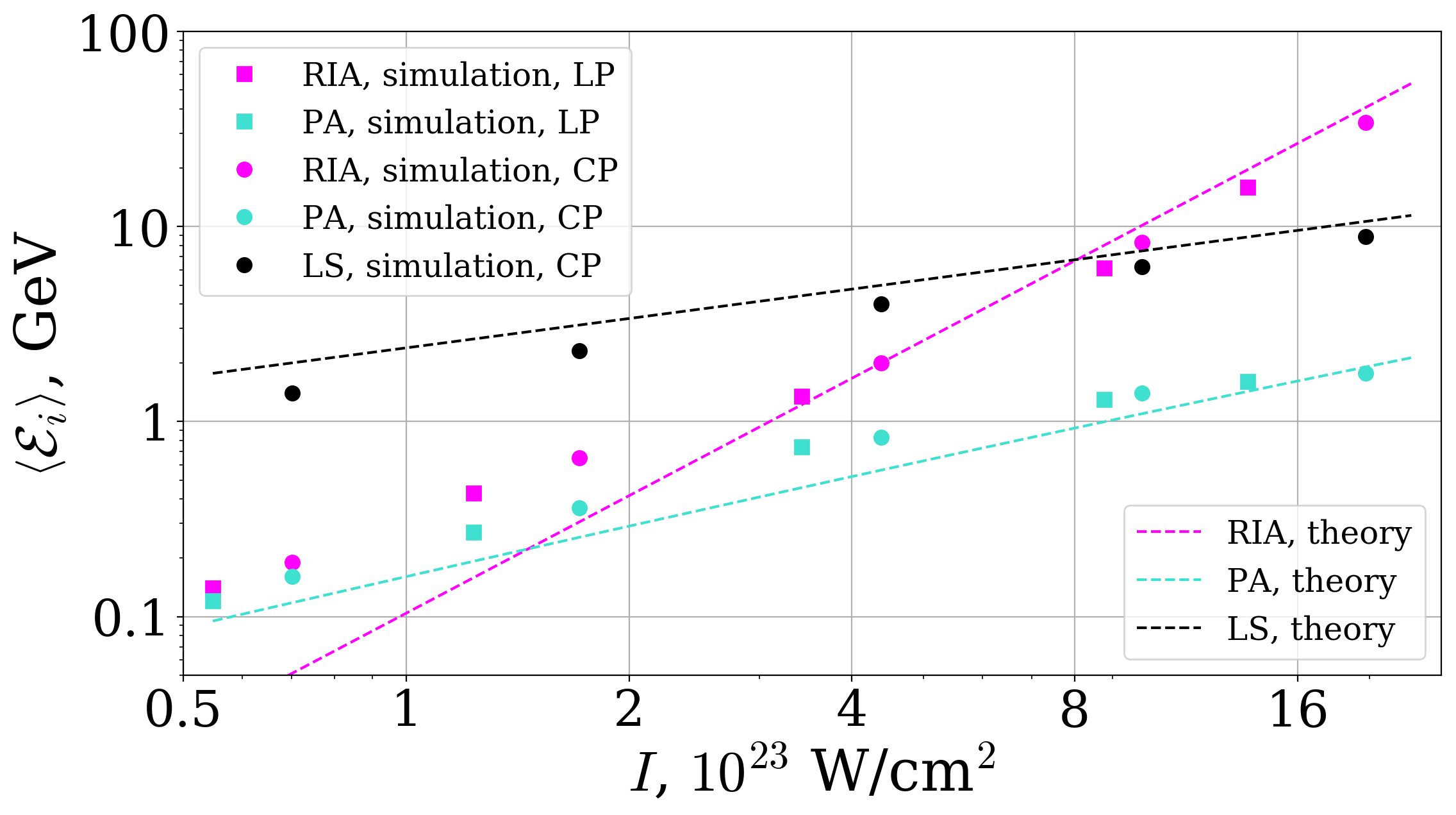}}}{0.55cm}{3.5cm}
\caption{\label{fig2}Average ion energies from 1D numerical simulations for different acceleration mechanisms. (a): average ion energy vs. time for a CP laser pulse with $a_0=600$ ($I\approx 10^{24}$ W/cm$^2$); inset: ion spectra at $t=25$ ps. (b): average ion energy at $t=25$ ps vs. laser intensity. Target density $n=0.5 n_c$, thickness $d=\lambda=1\mu m$; laser pulse FWHM $30$ fs. }
\end{figure}

As shown in Figure~\ref{fig2}, the resulting estimates (\ref{eavria}) -- (\ref{eavlim}) are in good agreement with 1D PIC simulations. The results for linear polarization (LP) can be obtained by the substitution \cite{pw_ppcf} $a_0\to a_0/\sqrt{2}$. One can also see that for strong ($I\gtrsim 10^{23}$ W/cm$^2$) laser pulses RIA is much more efficient than PA due to a stronger scaling with $a_0$. Comparing (\ref{eavpa}) and (\ref{eavlim}) we again arrive at the condition (\ref{rfdom}) for the dominance of RF. In particular, for low intensities (i.e. if (\ref{rfdom}) is violated), the ponderomotive force is stronger than the RF induced longitudinal force, and simulations with RF taken into account follow the estimate (\ref{eavpa}) for PA, see Figure~\ref{fig2}(b). 



Note that RIA can outperform (in terms of the mean ion energy) LS acceleration mechanism, commonly accepted as favorable for intense lasers and almost unaffected by RF \cite{tamburini2010}, see Figure~\ref{fig2}(b).  Let us discuss this point in more detail. According to a convential 1D model for LS acceleration \cite{esirkepov2004}, the energy of the laser pulse is almost entirely converted into the energy of the ions. By taking into account that the optimal foil areal density for LS is $\sigma_0\sim 2a_0$  \cite{macchi2009}, this is equivalent to the scaling
\begin{equation}\label{LS_scaling}
\left<\mathcal{E}_i^{\rm (LS)}\right>\sim m_e c^2 \frac{a_0^2 T}{\sigma_0}\sim m_e c^2 \frac{a_0 T}{2}.
\end{equation}
Then, by comparing (\ref{LS_scaling}) to (\ref{eavlim}), we conclude that RIA can outperform LS for 
\begin{equation}
\mu a_0^3\gtrsim \frac{4}{3},
\end{equation}
i.e. for $a_0\gtrsim 500$, or $I\gtrsim 7\cdot 10^{23}$ W/cm${}^2$, if $\lambda=1\mu m$. It very well agrees with numerical simulations,  see Figure~\ref{fig2}(b). Note that, in contrast to LS, in the RIA case only a small fraction of laser energy is transferred to particles (as is seen, e.g., from that the Gaussian laser profile in Figure~\ref{fig1}(a) remains undistorted). Nevertheless, being distributed among much fewer number of particles (as compared to LS), this can still provide higher average energy.


\begin{figure}[ht!]
\topinset{(a)}{\subfloat{\includegraphics[width=0.5\linewidth]{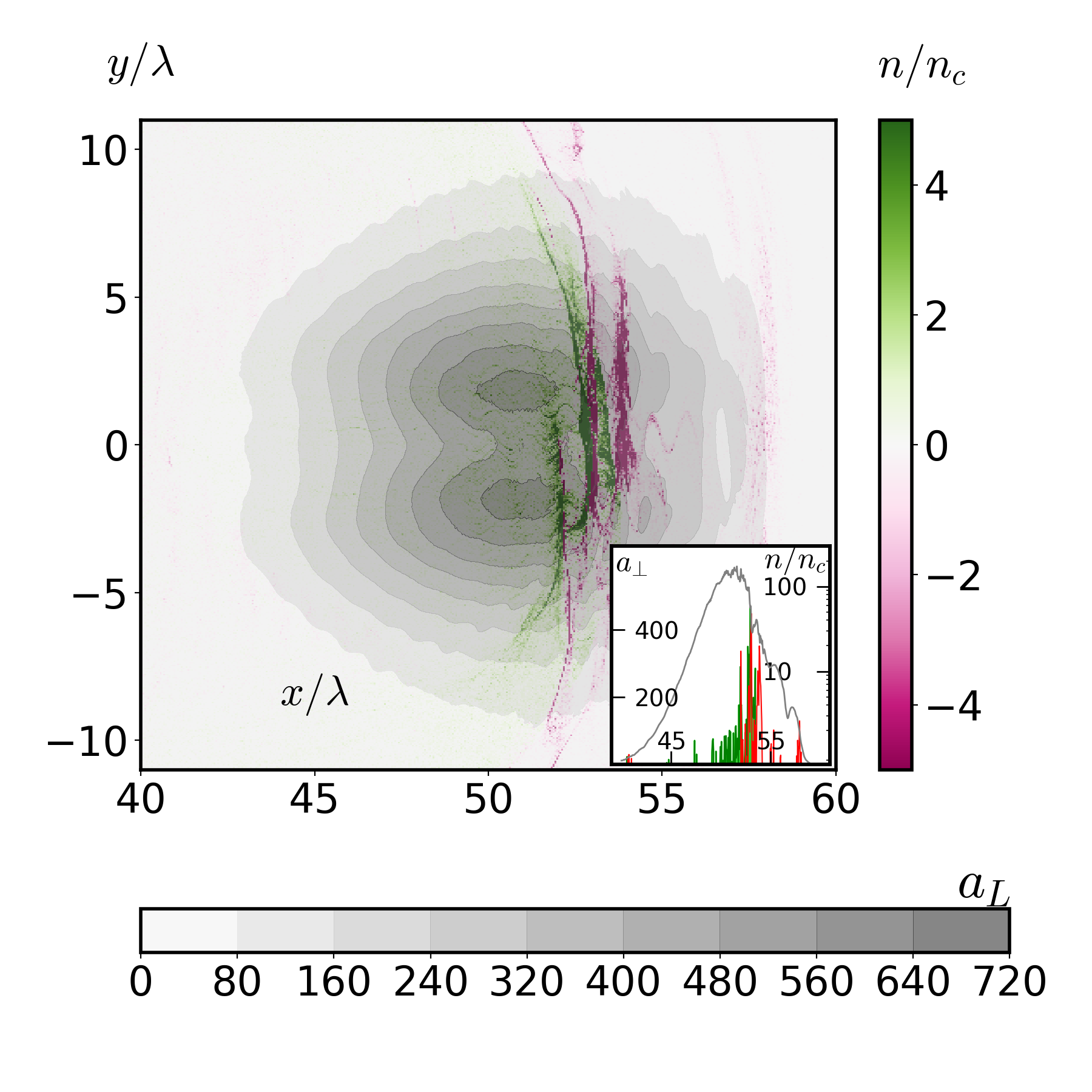}}}{0.95cm}{1.2cm}
\topinset{(b)}{\subfloat{\includegraphics[width=0.5\linewidth]{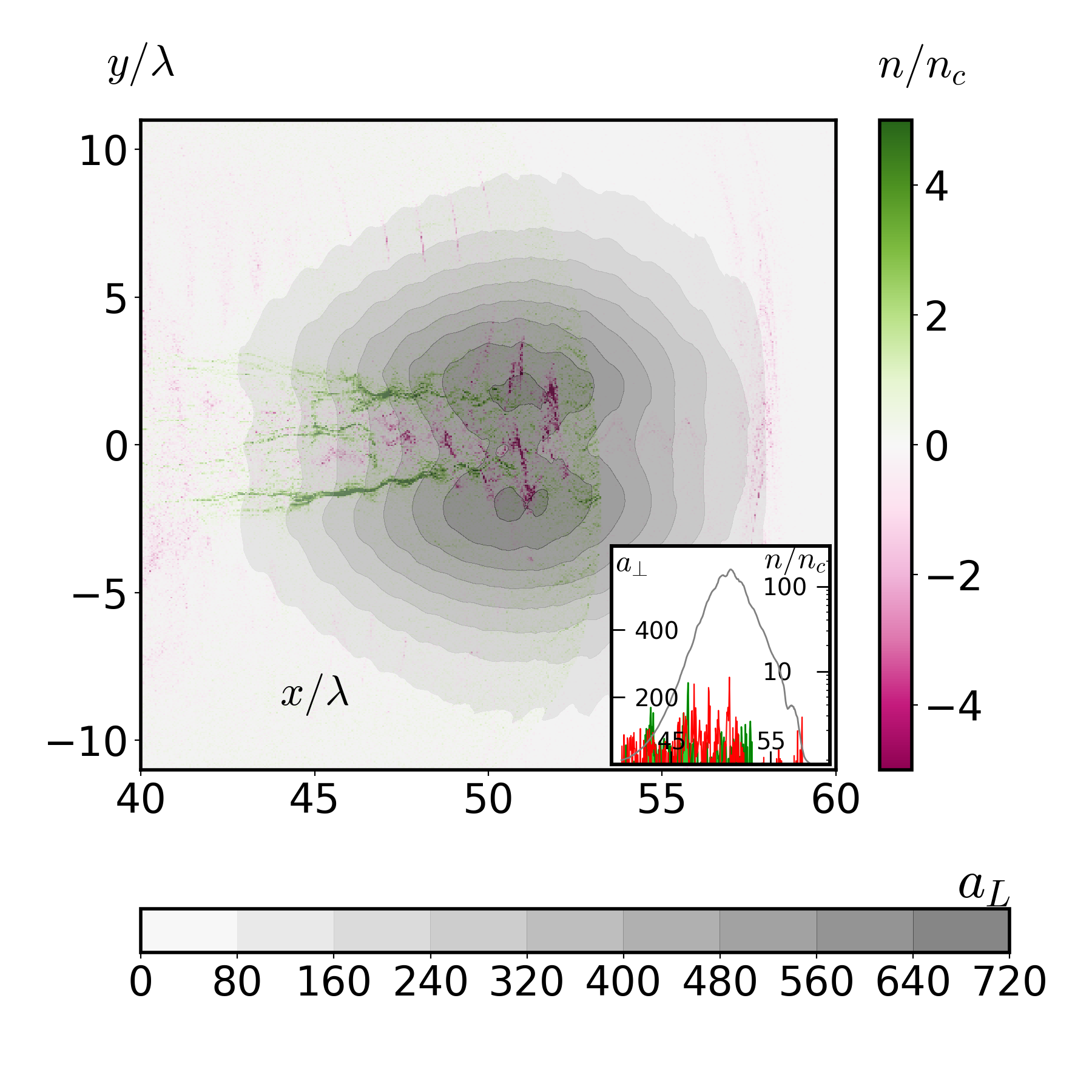}}}{1.3cm}{1.2cm}
\caption{\label{fig3} 2D PIC simulation results for ion acceleration in a transparent target: charge density (color scale) and laser field (grey scale) distributions at $t=127$ fs with (a) and without (b) RF; insets: sectional view along $y=0$. Laser pulse parameters: $I=1.3\cdot 10^{24}$ W/cm$^2$, FWHM pulse duration $30$ fs, waist radius $w=6\lambda$, circular polarization; target density $n_0=16n_c$ and thickness $d=\lambda$.}
\end{figure}

\begin{figure}[ht!]
\begin{center}
\includegraphics[width=0.5\linewidth]{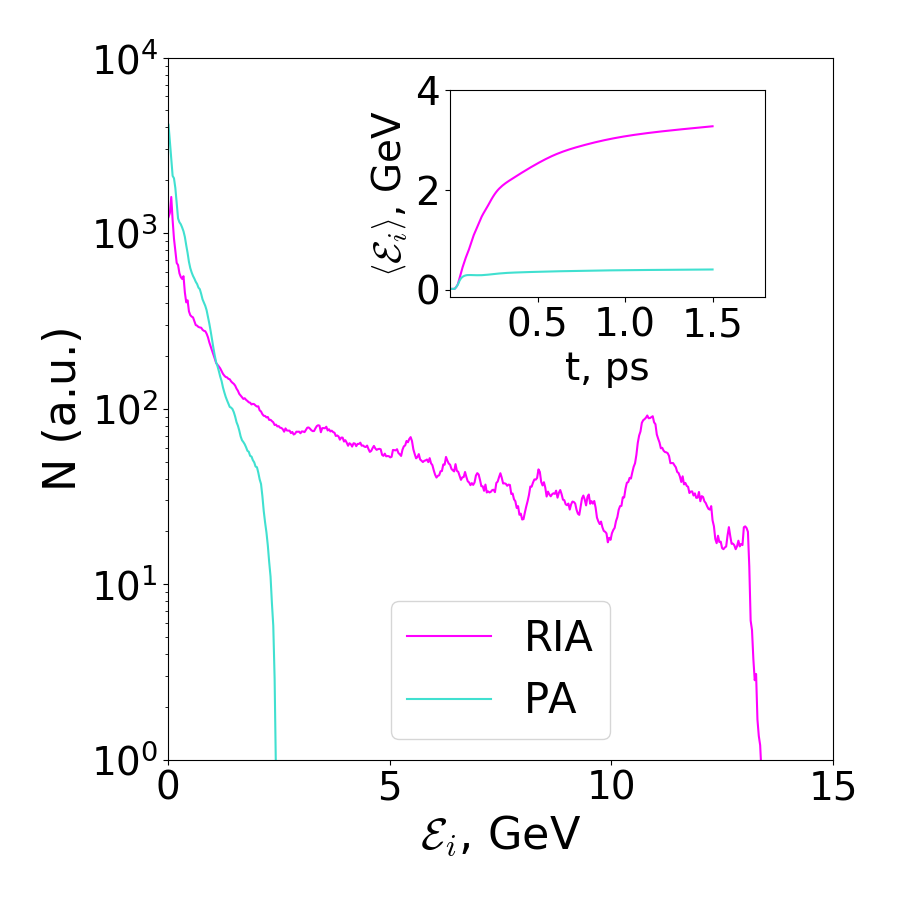}
\caption{\label{fig_spect2d} Ion spectra for RIA and PA at  $t=1440$ fs; inset: average ion energy for RIA and PA vs time. Laser and target parameters are similar to Fig.~\ref{fig3}. }
\end{center}
\end{figure}

In order to test RIA under more realistic conditions, we also performed 2D simulations with a focused laser pulse, see Figure~\ref{fig3}. Here, the ponderomotive force expels the electrons in transverse direction, thus reducing the charge separation field  \cite{pw_scirep}, see Figures~\ref{fig3}(a,b). Besides, the laser field amplitude reduces with time due to diffraction of the pulse. These negative effects  for RIA are partially compensated by optimizing the value of the target charge areal density $\sigma_0$ to be higher than in 1D case. This reduces the breakdown time $\tau_{bd}$ with respect to the transverse expansion and laser diffraction times. As a result, RF still allows more electrons to propagate inside the pulse (compare Figures~\ref{fig3}(a) and (b)), thus increasing the ions energy, as well as the total number of accelerated ions, see Figure~\ref{fig_spect2d}. However, the estimates (\ref{umaxria}), (\ref{umaxpa}) and (\ref{eavria})--(\ref{eavlim}) are no more accurate quantitatively and should be refined by taking correctly into account  the 2D effects. 

\section{Conclusions}

We examined the impact of RF on ion acceleration by strong laser pulses in a thin transparent foil. For each of the alternative acceleration mechanisms (RIA and PA, based on pushing the electrons forward either by the RF induced longitudinal Lorentz force, or by the ponderomotive force, respectively), we elaborated a proper 1D analytical model. As shown in Figures~\ref{fig_maxen}(a) and \ref{fig2}, each model is in excellent agreement with 1D PIC simulations. By using either (\ref{eavria}) -- (\ref{eavlim}) or 1D PIC simulations, we identified the parameter regions where RF essentially enhances ion acceleration and RF induced acceleration can be favorable among other acceleration mechanisms. The effect remains valid in multi dimensions, however, its accurate quantitative description is a difficult separate problem still to be solved.

\appendix

\section{Limits of applicability of the RIA model}
Let us briefly review the assumptions, which we made for consideration of electrons dynamics in the RIA case. 
\begin{itemize}
\item First, we assumed that electrons motion is ultrarelativistic in longitudinal direction, and hence $u_x\gg a$. Using (\ref{urf}) with $\sigma\sim\sigma_0$ and $a\sim a_0$ as above, we get
\begin{equation}\label{sepcond}
\mu a_0^4\gg\sigma_0.
\end{equation}
\item Second, we assumed that $u_\bot\approx a_0$, i.e. the transverse component of the RF force is much smaller than the Lorentz force \cite{pw_scirep,pw_ppcf}. It means that $\mu a_0^2\gamma^2(1-v_x)v_\bot\ll a_0$,
or, using $a_0\ll u_x$ and (\ref{urf}) with $\sigma\sim\sigma_0$,
\begin{equation}
\mu a_0^2\sigma_0\ll 1.
\end{equation}
\item Third, we assumed that the first and second terms in Landau-Lifshitz expression \cite{LL} for the RF force
\begin{equation}
\eqalign{
\mathbf{F}_{RF}=\frac{2e^3}{3mc^3}\gamma\left\{\left(\frac{\partial}{\partial t}+(\mathbf{v}\nabla)\right)\mathbf{E}+\frac{1}{c}\left[\mathbf{v}\times\left(\frac{\partial}{\partial t}+(\mathbf{v}\nabla)\right)\mathbf{B}\right]\right\}-\cr
-\frac{2e^4}{3m^2c^4}\left\{[\mathbf{E}\times\mathbf{B}]+\frac{1}{c}[\mathbf{B}\times[\mathbf{B}\times\mathbf{v}]]+\frac{1}{c}\mathbf{E}(\mathbf{v}\mathbf{E})\right\}-\cr
-\frac{2e^4}{3m^2c^5}\gamma^2\mathbf{v}\left\{\left(\mathbf{E}+\frac{1}{c}[\mathbf{v}\times\mathbf{B}]\right)^2-\frac{1}{c^2}(\mathbf{Ev})^2\right\}}
\end{equation}
are much smaller than the third term, which we solely substituted into Equation~(\ref{lleq}). Indeed, contributions from the first in second terms in the dimensionless form can be written as
\begin{equation}
\mathbf{f}^{RF}_{1+2}=\mu(1-v_x)\left[\gamma\mathbf{a}(1-v_x)+a^2\right],
\end{equation}
and it is $\sim a^2\gg1$ times smaller than the RF-induced term in the Equation~(\ref{eq1p}).
\item Finally, we used the classical approximation for RF, which implies that the quantum parameter (\ref{chi}) $\chi\ll 1$ \cite{ritus1985,dipiazza2012}. In the ultrarelativistic limit under consideration 
\begin{equation}
\chi\sim \frac{\mu}{\alpha}\gamma a_0(1-v_x)\sim \frac{\sqrt{\sigma_0\mu}}{\alpha}\ll1.
\end{equation}
\end{itemize}

\section{Numerical approach}
For numerical simulations we used the PIC code SMILEI  \cite{SMILEI}, which allows to make calculations with and without RF taken into account. In all simulations we used laser pulses with the wavelength $\lambda=1\mu$m and the Gaussian temporal profile $a(\varphi)=a_0\exp[-(\varphi-4T)^2/T^2]$. The dimensionless  pulse duration $T$ is related to FWHM duration by $T\approx 0.57\omega t_{\rm FWHM}$. For 2D simulations we used laser pulses with a supergaussian spatial profile, see  \cite{smilei_pulses2018} for the details of implementation of arbitrary pulse shapes in SMILEI. The intensity distribution in a focal plane was of the form $I\sim\exp[-(y/w)^8]$, where the laser pulse waist $w=6\lambda$.

For 1D simulations we used $1000$ cells per wavelength to resolve the limit $\Delta x<\lambda/a_0$, established in \cite{arefiev_pop2015}. Note that \cite{arefiev_pop2015} does not take RF into account, however, we checked that results are similar for $\Delta x=\lambda/1000$ and $\Delta x=\lambda/500$. We took $50$ particles of each type per cell for low density foils and $n/n_c$ particles per cell for high density ones; time step was equal to a space step divided by the factor $c/0.95$. 2D simulations were done inside a $500\lambda\times 128\lambda$ box with the cell side length $dx=dy=\lambda/100$ in both directions. Time step was chosen as $dt=0.95dx/c\sqrt{2}$, and the number of particles of each type per cell was equal to $16$. Target density for LS simulations satisfied the condition  \cite{macchi2009} $l/\lambda=a_0/\pi\tilde{n}$. Boundary conditions for particles and fields were absorbing. The target thickness was $d=\lambda$, and the laser was focused on its left boundary, located at $x=30\lambda$.

\ack
We are grateful to S.V. Bulanov, S.S. Bulanov, O. Klimo, T. Schlegel, M. Vranic for valuable discussions, and to M. Grech for both valuable discussions and assistance with the code.
The research was performed using the code SMILEI  and the resources of the ELI Beamlines Eclipse cluster, and was supported by the projects ADONIS (Advanced research using high intensity laser produced photons and particles) 
CZ.02.1.01/0.0/0.0/16\_019/0000789 and HiFI (High-Field Initiative) CZ.02.1.01/0.0/0.0/15003/0000449 from European Regional Development Fund, the MEPhI Academic Excellence Project (Contract No. 02.a03.21.0005), the Russian Foundation for Basic Research (Grant 19-02-00643), and the Tomsk State University Competitiveness Improvement Program.

\end{document}